\documentclass[twocolumn,aps,pra,amsmath,amssymb]{revtex4-1}
\usepackage[latin9]{inputenc}
\usepackage{amsmath}
\usepackage{amssymb}
\usepackage{graphicx}
\usepackage{esint}

\makeatletter

\usepackage{epsfig}
\usepackage{bm}
\usepackage{dcolumn}
\usepackage{color}

\makeatother

\begin{document}
\title{Raman spectroscopy of Fermi polarons}
\author{Hui Hu$^{1}$ and Xia-Ji Liu$^{1}$}
\affiliation{$^{1}$Centre for Quantum Technology Theory, Swinburne University
of Technology, Melbourne, Victoria 3122, Australia}
\date{\today}
\begin{abstract}
By using a non-self-consistent many-body $T$-matrix theory, we calculate
the finite-temperature Raman spectroscopy of a mobile impurity immersed
in a Fermi bath in three dimensions. The dependences of the Raman
spectrum on the transferred momentum, temperature, and impurity-bath
interaction are discussed in detail. We confirm that the peak in the
Raman spectrum shows a weaker dependence on the impurity concentration
than that in the radio-frequency spectroscopy, due to the nonzero
transferred momentum, as anticipated. We compare our theoretical prediction
with the recent measurement by Gal Ness \textsl{et al.} in Physical
Review X \textbf{10}, 041019 (2020) without any adjustable parameters.
At weak coupling, we find a good quantitative agreement. However,
close to the Feshbach resonance the agreement becomes worse. At strong
coupling, we find that an unrealistic Fermi bath temperature might
be needed, in order to account for the experimental data.
\end{abstract}
\maketitle

\section{Introduction}

Fermi polarons - quasiparticles formed when impurities move inside
a Fermi bath - have received increasing attentions from researchers
in different research fields, due to the rapid experimental advances
in ultracold atomic physics. Nowadays, Fermi polarons can be routinely
realized by using a highly imbalanced two-component Fermi-Fermi mixture
\cite{Massignan2014,Lan2014,Schmidt2018,Wang2022Review}, where the
minority atoms play the role of impurities. The interaction between
the impurity and the majority Fermi bath or Fermi sea can be tuned
precisely with the help of Feshbach resonances \cite{Chin2010}. Many
useful techniques have been developed to characterize the quasiparticle
properties of Fermi polarons, such as the radio-frequency spectroscopy
\cite{Schirotzek2009,Zhang2012,Kohstall2012,Koschorreck2012,Scazza2017,Zan2019},
Ramsey interferometry \cite{Cetina2016}, Rabi oscillation \cite{Kohstall2012,Scazza2017},
and Raman spectroscopy \cite{Ness2020}. As a result, a number of
intriguing features of Fermi polarons have been revealed \cite{Wang2022PRL,Wang2022PRA},
including the excited branch of repulsive polarons \cite{Kohstall2012,Koschorreck2012,Scazza2017,Massignan2011},
and the disappearance of attractive polarons at sufficiently large
impurity-bath coupling \cite{Ness2020,Prokofev2008,Punk2009,Combescot2009}.

Here, we are interested in the Raman spectroscopy, which has been
applied most recently by Gal Ness and co-workers to observe the polaron-molecule
transition \cite{Ness2020}. In this experiment, the Fermi bath temperature
is about one-fifth Fermi temperature (i.e., $0.2T_{F}$), and the
trap-average impurity concentration, defined by $x=n_{\textrm{imp}}/n$
with $n_{\textrm{imp}}$ ($n$) being the density of impurity and
(bath) atoms, is about $0.23$. To account for the temperature effect,
the experimental data have been analyzed by using a \emph{phenomenological}
theory, where the system is treated as a non-interacting thermal mixture
of polarons and molecules \cite{Ness2020}. The thermal distributions
or numbers of polarons and molecules are determined respectively according
to their zero-temperature energy, calculated using the variational
Chevy ansatz at the lowest level of particle-hole excitations \cite{Chevy2006}.
A set of free fitting parameters, such as the polaron residue $\mathcal{Z}$,
polaron temperature $T_{P}$ (which might be different from the Fermi
bath temperature) and effective binding energy of molecules, have
been introduced to fit the data and to extract the polaron energy
and molecule binding energy \cite{Ness2020}.

In this work, we would like to theoretically understand the measured
Raman spectrum, based on a finite-temperature \emph{microscopic} theory
within the non-self-consistent many-body $T$-matrix approximation
\cite{Hu2022,hu2022CrossoverPolaron}. At zero temperature, the non-self-consistent
$T$-matrix approximation is fully equivalent to the variational Chevy
ansatz with one-particle-hole excitation \cite{Combescot2007}. The
finite-temperature extension of the $T$-matrix theory \cite{Hu2022,Hu2018,Mulkerin2019}
allows us to improve the phenomenological treatment used in the experimental
analysis, at least in the weak coupling regime, where attractive polarons
are well-defined quasiparticles in the ground state \cite{Punk2009,Combescot2007}.

Indeed, for weak coupling we find that our theory agrees very well
with the experimental data, without any free adjustable parameters.
In the vicinity of the Feshbach resonance of the impurity-bath interaction,
however, the experimental data appear to lie systematically lower
than the theoretical prediction. To theoretically account for the
data, we need to significantly increase the Fermi bath temperature.
This is somehow consistent with the experimental observation that
the extracted polaron temperature $T_{P}$ is systematically higher
than the Fermi bath temperature near the Feshbach resonance. Finally,
for the strong coupling case our theory fails to explain the experimental
data with any reasonable Fermi bath temperature. This failure clearly
indicates the importance of developing a better description of Fermi
polarons near the polaron-molecule transition, by extending more reliable
zero-temperature theories such as the functional renormalization group
\cite{Schmidt2011} and diagrammatic Monte Carlo simulation \cite{Prokofev2008,Goulko2016}
to the finite-temperature case. 

The rest of the paper is organized as follows. In the next section
(Sec. II), we briefly review the non-self-consistent many-body $T$-matrix
theory for Fermi polarons at finite temperature. In Sec. III, we first
discuss in detail the properties of Raman spectroscopy in the ejection
scheme and investigate the spectrum as functions of the transferred
momentum, impurity concentration, temperature, and the impurity-bath
interaction strength. We then compare our theoretical predictions
with the experimental data. We also briefly consider the injection
Raman spectrum. Finally, Sec. IV is devoted to the conclusions and
outlooks. In Appendix A, we also consider the effect of spatial inhomogeneity
on the Raman spectrum, caused by the external harmonic traps.

\section{The non-self-consistent many-body $T$-matrix theory}

The non-self-consistent many-body $T$-matrix theory of Fermi polarons
at finite temperature has been discussed at length in Ref. \cite{Hu2022}.
Here, for self-containedness we briefly review the key equations in
the following. The Fermi polaron system under consideration involves
an impurity of mass $m_{I}$ interacting with a homogeneous bath of
fermionic atoms of mass $m$. It can be described by a model Hamiltonian,

\begin{equation}
\mathcal{H}=\sum_{\mathbf{k}}\epsilon_{\mathbf{k}}c_{\mathbf{k}}^{\dagger}c_{\mathbf{k}}+\sum_{\mathbf{p}}\epsilon_{\mathbf{p}}^{(I)}d_{\mathbf{p}}^{\dagger}d_{\mathbf{p}}+\frac{g}{V}\sum_{\mathbf{kpq}}c_{\mathbf{k}}^{\dagger}d_{\mathbf{q}-\mathbf{k}}^{\dagger}d_{\mathbf{\mathbf{q}-p}}c_{\mathbf{p}},
\end{equation}
where $c_{\mathbf{k}}^{\dagger}$ ($d_{\mathbf{p}}^{\dagger}$) are
the creation field operators for fermionic atoms (impurity) with momentum
$\mathbf{k}$ ($\mathbf{p}$) and single-particle dispersion relation
$\epsilon_{\mathbf{k}}=\hbar^{2}\mathbf{k}^{2}/(2m)$ ($\epsilon_{\mathbf{p}}^{(I)}=\hbar^{2}\mathbf{p}^{2}/(2m_{I})$),
and $V$ is the system volume. The last term in the Hamiltonian describes
the $s$-wave contact interaction between impurity and bath with a
bare coupling strength $g$, which is to be regularized via the relation,
\begin{equation}
\frac{1}{g}=\frac{m_{r}}{2\pi\hbar^{2}a}-\frac{1}{V}\sum_{\mathbf{p}}\frac{2m_{r}}{\hbar^{2}\mathbf{p}^{2}}.
\end{equation}
Here, $a$ is the $s$-wave scattering length between impurity and
the Fermi bath, $m_{r}\equiv mm_{I}/(m+m_{I})$ is the reduced mass
for the impurity-bath scattering. Throughout the work, we always take
$m_{I}=m$, so $m_{r}=m/2$. The number of fermionic atoms in the
Fermi bath ($n$) can be tuned by adjusting the chemical potential
$\mu(T)$ at nonzero temperature $T$. We often measure the single-particle
energy of the bath from the chemical potential and therefore define
$\xi_{\mathbf{k}}\equiv\epsilon_{\mathbf{k}}-\mu$.

To solve the Fermi polaron problem, the key quantity of interest is
the (retarded) impurity Green function,

\begin{equation}
G_{R}\left(\mathbf{k},\omega\right)=\frac{1}{\omega-\epsilon_{\mathbf{k}}^{(I)}-\Sigma_{R}\left(\mathbf{k},\omega\right)},\label{eq:impurityGF}
\end{equation}
and its associated single-particle spectral function, $A(\mathbf{k},\omega)\equiv-(1/\pi)\textrm{Im}G_{R}(\mathbf{k},\omega)$.
In the single-impurity limit, within the non-self-consistent many-body
$T$-matrix theory \cite{Hu2022}, the (retarded) self-energy $\Sigma_{R}(\mathbf{k},\omega)$
can be calculated by summing all the ladder-type diagrams, which gives
rise to an expression,
\begin{equation}
\Sigma_{R}\left(\mathbf{k},\omega\right)=\frac{1}{V}\sum_{\mathbf{q}}f\left(\xi_{\mathbf{q}-\mathbf{k}}\right)\Gamma_{R}\left(\mathbf{q},\omega+\xi_{\mathbf{q}-\mathbf{k}}\right),\label{eq:selfenergy}
\end{equation}
where $f(x)\equiv1/(e^{\beta x}+1)$ with $\beta\equiv1/(k_{B}T)$
being the Fermi-Dirac distribution function, and $\Gamma_{R}(\mathbf{q},\Omega)$
is the vertex function, whose inverse is given by ($\Omega^{+}\equiv\Omega+i0^{+}$),
\begin{eqnarray}
\chi_{R}(\mathbf{q},\Omega) & \equiv & \Gamma_{R}^{-1}(\mathbf{q},\Omega)=\chi_{R}^{(2b)}+\chi_{R}^{(mb)},\\
\chi_{R}^{(2b)} & = & \frac{m_{r}}{2\pi\hbar^{2}a}-\frac{1}{V}\sum_{\mathbf{k}}\left[\frac{1}{\Omega^{+}-\xi_{\mathbf{k}}-\epsilon_{\mathbf{q}-\mathbf{k}}^{(I)}}+\frac{2m_{r}}{\hbar^{2}\mathbf{k}^{2}}\right],\nonumber \\
\chi_{R}^{(mb)} & = & \frac{1}{V}\sum_{\mathbf{k}}\frac{f\left(\xi_{\mathbf{k}}\right)}{\Omega^{+}-\xi_{\mathbf{k}}-\epsilon_{\mathbf{q}-\mathbf{k}}^{(I)}}.\nonumber 
\end{eqnarray}
As discussed in Ref. \cite{Hu2022}, the two-dimensional integrals
in the calculations of the inverse vertex function $\Gamma_{R}^{-1}(\mathbf{q},\Omega)$
can be efficiently evaluated, leading to fast and accurate determination
of the (retarded) self-energy $\Sigma_{R}(\mathbf{k},\omega)$ and
consequently the single-particle spectral function $A(\mathbf{k},\omega)$.

\begin{figure}
\begin{centering}
\includegraphics[clip,width=0.48\textwidth]{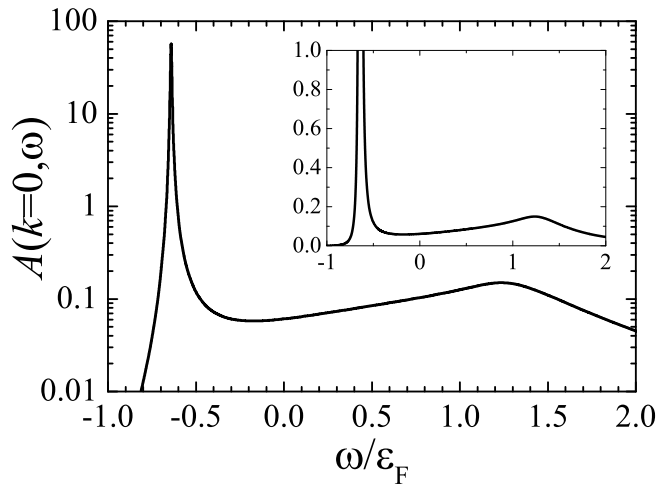}
\par\end{centering}
\caption{\label{fig1_akw} The zero-momentum impurity spectral function $A(\mathbf{k}=0,\omega$)
in the unitary limit $1/a=0$. The spectral function is in units of
of $\varepsilon_{F}^{-1}$, where $\varepsilon_{F}\equiv\hbar^{2}k_{F}^{2}/(2m)$
and $k_{F}=(6\pi^{2}n)^{1/3}$ are the Fermi energy and Fermi wavevector,
respectively. The temperature is $T=0.2T_{F}=0.2\varepsilon_{F}/k_{B}$.
The inset shows the spectral function in the linear scale.}
\end{figure}

As an example, in Fig. \ref{fig1_akw} we show the zero-momentum impurity
spectral function $A(\mathbf{k}=0,\omega$) at the Feshbach resonance,
where the $s$-wave scattering length $a$ between the impurity and
Fermi bath diverges. A finite temperature typically leads to a nonzero
thermal decay of the attractive Fermi polaron, as given by $\Gamma=-2\mathcal{Z}\textrm{Im}\Sigma_{R}(\mathbf{0},\mathcal{E}_{P})$,
where $\mathcal{E}_{P}$ is the polaron energy and $\mathcal{Z}=[1-\partial\textrm{Re}\Sigma_{R}(0,\omega)/\partial\omega]_{\omega=\mathcal{E}_{P}}^{-1}$
is the residue of the attractive polaron. At the low temperature considered
in Fig. 1 (i.e., $T=0.2T_{F}$), the decay rate $\Gamma\simeq0.008\varepsilon_{F}$
is very small, as indicated by the narrow full width at half maximum
(FWHM) of the spectral function. This leads to a sharp peak at the
polaron energy $\mathcal{E}_{P}\simeq-0.64\varepsilon_{F}$. There
is also a broad peak at much higher energy $\sim\varepsilon_{F}$,
which may be understood as a precursor of the repulsive polaron. Our
accurate determination of the impurity spectral function is able to
capture both sharply peaked attractive polaron and broadly distributed
repulsive polaron, allowing us to perform a microscopic calculation
of the Raman spectrum at finite temperature, as we shall discuss in
detail as follows.

\section{Results and discussions}

\begin{figure}
\begin{centering}
\includegraphics[width=0.48\textwidth]{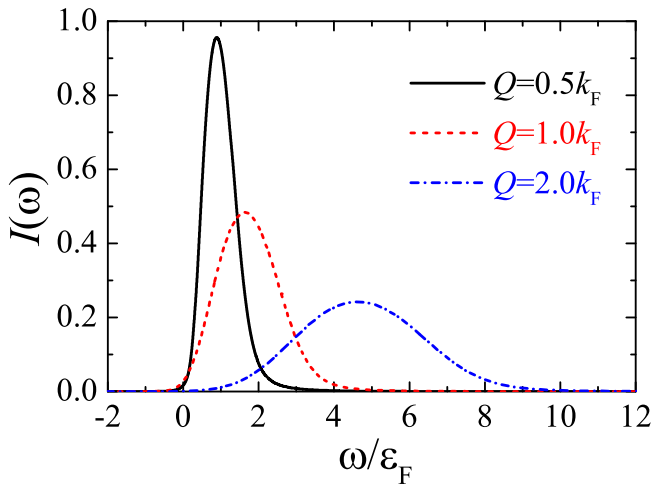}
\par\end{centering}
\caption{\label{fig2_ejRaman_Qdep} The ejection Raman spectrum in the unitary
limit and $T=0.2T_{F}$, at different transferred momenta as indicated.
The spectra are in units of $\varepsilon_{F}^{-1}$, and are normalized
to unity (i.e., $\int d\omega I(\omega)=1$). This can be achieved
by dividing $I(\omega)$ the impurity density $n_{\textrm{imp}}$.
We have taken an impurity concentration $n_{\textrm{imp}}=0.15n$.}
\end{figure}

\subsection{Ejection Raman spectroscopy}

In the Raman spectroscopy experiment \cite{Ness2020}, initially the
impurity is in the interacting state with the Fermi bath. It is then
transferred to a non-interacting state using Raman beams with the
energy $\omega$ and momentum $\mathbf{Q}$. In this \emph{ejection}
scheme, according to the linear response theory the transfer rate
is proportional to \cite{Ness2020,Torma2014,Parish2021},

\begin{equation}
I\left(\omega\right)=\frac{1}{V}\sum_{\mathbf{k}}A\left[\mathbf{k},\epsilon_{\mathbf{k}+\mathbf{Q}}^{(I)}-\omega\right]f\left(\epsilon_{\mathbf{k}+\mathbf{Q}}^{(I)}-\omega-\mu_{I}\right).\label{eq:ejectionRaman}
\end{equation}
Here, we have assumed a fermionic impurity according to the experiment
\cite{Ness2020}. To account for the finite impurity density $n_{\textrm{imp}}$,
we have also introduced an impurity chemical potential $\mu_{I}$,
which is to be determined by the number equation,
\begin{equation}
n_{\textrm{imp}}=\frac{1}{V}\sum_{\mathbf{k}}\intop_{-\infty}^{+\infty}d\omega f\left(\omega-\mu_{I}\right)A\left(\mathbf{k},\omega\right).
\end{equation}
It is readily seen that the ejection Raman spectrum is normalized
to the impurity density, i.e., $\int d\omega I(\omega)=n_{\textrm{imp}}$.
Moreover, in the limit of zero transferred momentum $Q=0$, the Raman
spectroscopy simply recovers the well-studied radio-frequency spectroscopy
\cite{Schirotzek2009,Zan2019,Hu2022}. In Fig. \ref{fig2_ejRaman_Qdep},
we report exemplified Raman spectra at different transferred Raman
momenta for the unitary impurity-bath interaction. As the momentum
$Q$ increases, the Raman peak becomes broader and quickly shifts
to high energy.

\begin{figure}
\begin{centering}
\includegraphics[clip,width=0.48\textwidth]{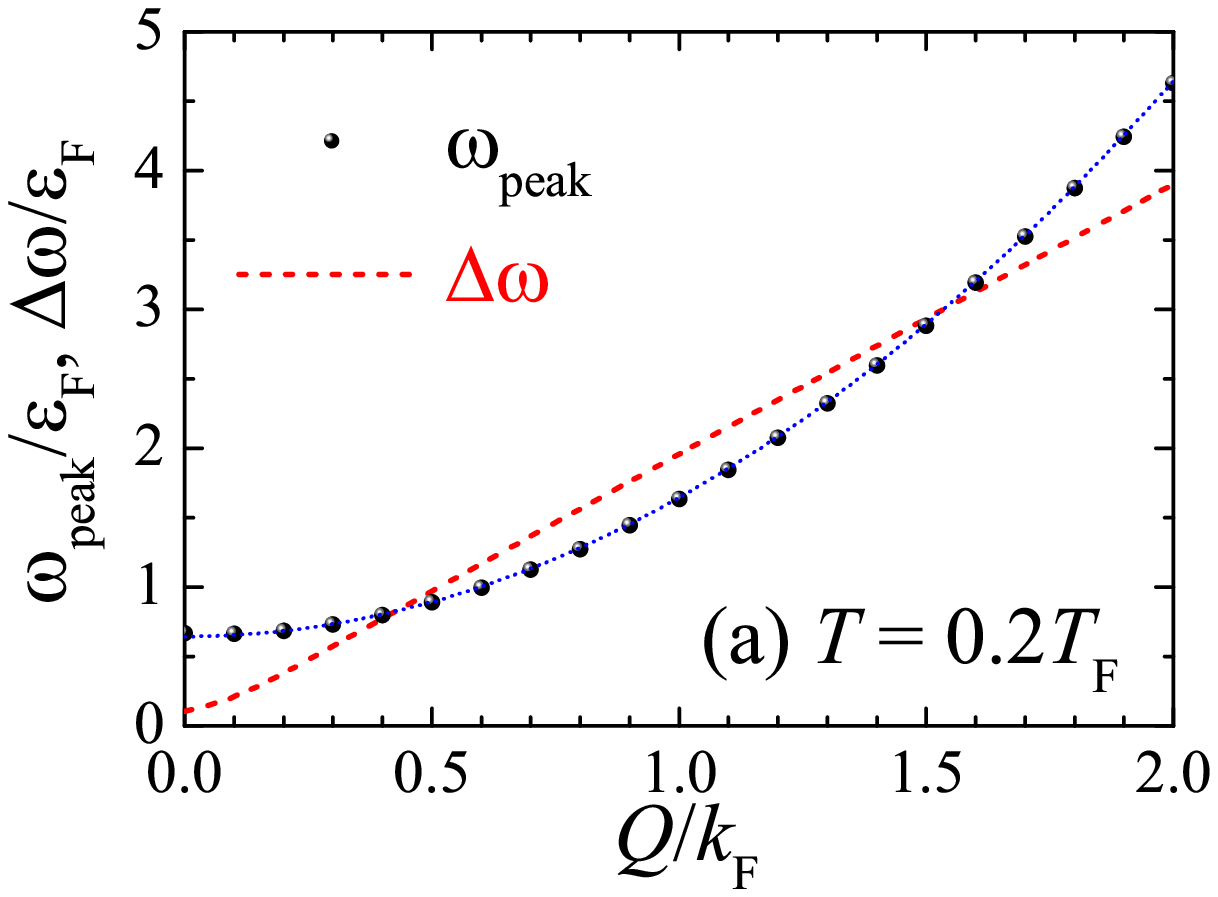}
\par\end{centering}
\begin{centering}
\includegraphics[clip,width=0.48\textwidth]{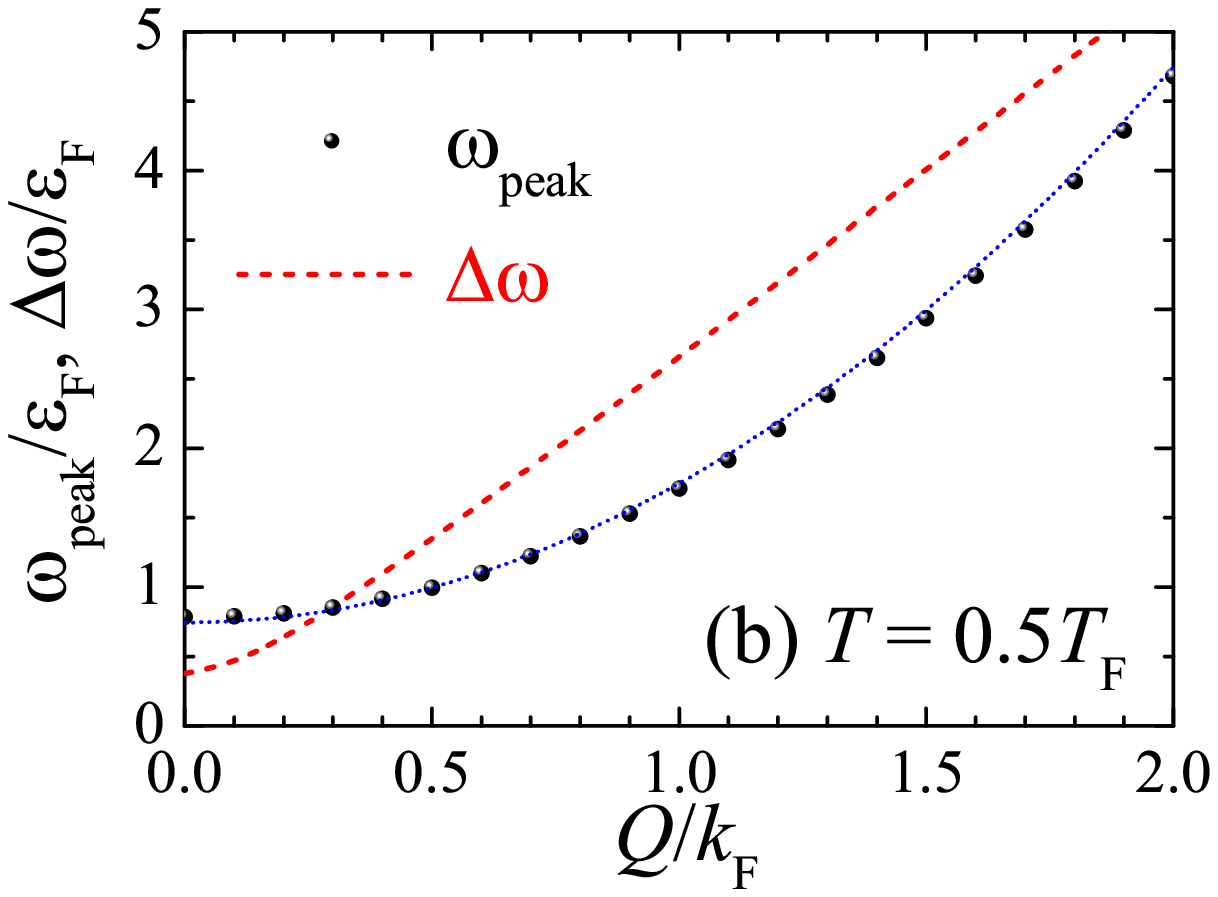}
\par\end{centering}
\caption{\label{fig3_ejRaman_w0Q} The position and width of the peak in the
ejection Raman spectrum as a function of the transferred momentum
$Q$, at $T=0.2T_{F}$ (a) and $T=0.5T_{F}$ (b). Here, we consider
the unitary limit with $1/(k_{F}a)=0$. The blue dashed line in each
plot shows the anticipated peak position $\omega_{\textrm{peak}}=\hbar^{2}Q^{2}/(2m)-\mathcal{E}_{P}$.
The impurity concentration is $n_{\textrm{imp}}/n=0.15$.}
\end{figure}

The blue shift of the peak with transferred momentum can be easily
understood from the coherent part of the polaron spectral function,
which at zero temperature takes the form \cite{Ness2020},
\begin{equation}
A_{\textrm{coh}}(\mathbf{k},\omega)\simeq\mathcal{Z}\delta\left[\omega-\left(\mathcal{E}_{P}+\frac{\hbar^{2}k^{2}}{2m^{*}}\right)\right],
\end{equation}
where the polaron mass $m^{*}$ is close to the mass of bare atoms,
$m^{*}\simeq m_{I}=m$, unless near the polaron-molecule transition.
By substituting $A_{\textrm{coh}}$ into the ejection expression Eq.
(\ref{eq:ejectionRaman}), we find a coherent contribution to the
Raman signal, if the frequency $\omega$ satisfies 
\begin{equation}
\omega=\epsilon_{\mathbf{k}+\mathbf{Q}}^{(I)}-\frac{\hbar^{2}k^{2}}{2m^{*}}-\mathcal{E}_{P}\simeq\frac{\hbar^{2}Q^{2}}{2m}-\mathcal{E}_{P}+\frac{\hbar^{2}\mathbf{k}\cdot\mathbf{Q}}{m}.\label{eq:wRaman}
\end{equation}
After performing the integration over the angle between $\mathbf{k}$
and $\mathbf{Q}$, the coherent contribution is therefore centered
around the peak position, $\omega_{\textrm{peak}}\simeq\hbar^{2}Q^{2}/(2m)-\mathcal{E}_{P}$.
In Fig. \ref{fig3_ejRaman_w0Q}, we examine the dependence of the
Raman peak on the transferred momentum $Q$ at two characteristic
temperatures $T=0.2T_{F}$ and $0.5T_{F}$ in the unitary limit. The
peak position extracted from the Raman spectrum (solid circles) follows
the anticipated trajectory $\hbar^{2}Q^{2}/(2m)-\mathcal{E}_{P}$
(blue dotted lines), with the difference barely observable in the
scale of the figure. 

In Fig. \ref{fig3_ejRaman_w0Q}, we also report the width of the Raman
peak as a function of the transferred momentum $Q$ (see red dashed
lines). The width seems to be proportional to the transferred momentum
at large $Q$. It is also strongly temperature dependent. Both the
temperature dependence and linear $Q$-dependence might be understood
from the last term in Eq. (\ref{eq:wRaman}), since the width is directly
related to the maximum value of $Qk$, which depends on both temperature
and the impurity concentration. Therefore, from the width of Raman
spectrum we can hardly extract useful information about the decay
rate of Fermi polarons.

\begin{figure}[t]
\begin{centering}
\includegraphics[width=0.48\textwidth]{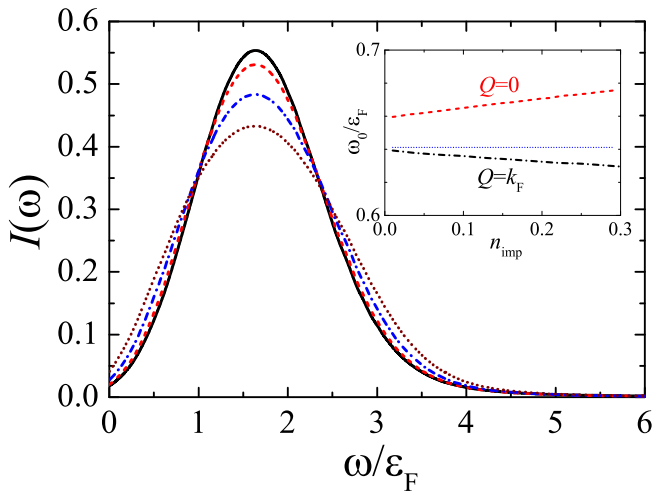}
\par\end{centering}
\caption{\label{fig4_ejRaman_nimp} The ejection Raman spectrum at different
impurity densities: $n_{\textrm{imp}}=0.01$ (black solid line), $0.05$
(red dashed line), $0.15$ (blue dot-dashed line), and $0.30$ (brown
dotted line), in units of the Fermi bath density $n$. Here, we consider
the unitary limit at $T=0.2T_{F}$. The transferred momentum is $Q=k_{F}$.
The spectra are in units of $\varepsilon_{F}^{-1}$, and are normalized
to unity (i.e., $\int d\omega I(\omega)=1$). The inset shows $\omega_{0}=\omega_{\textrm{peak}}-\hbar^{2}Q^{2}/(2m)$
for the radio-frequency spectrum (with $Q=0$) and for the Raman spectrum
(with $Q=k_{F}$). The blue dotted line is the anticipated polaron
energy $\omega_{0}=\left|\mathcal{E}_{P}\right|$ .}
\end{figure}

\begin{figure}[t]
\begin{centering}
\includegraphics[width=0.48\textwidth]{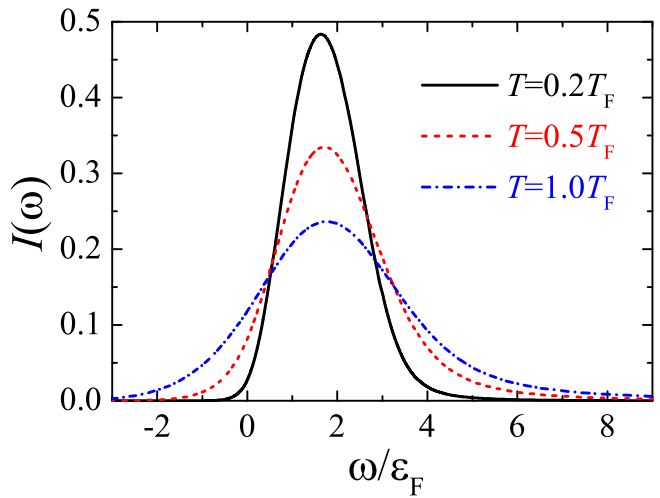}
\par\end{centering}
\caption{\label{fig5_ejRaman_Tdep} The temperature dependence of the ejection
Raman spectrum $I(\omega$) {[}in units of $\varepsilon_{F}^{-1}${]}
at three temperatures: $T/T_{F}=0.2$ (black solid line), $0.5$ (red
dashed line), and $1.0$ (blue dot-dashed line). Here, we take a transferred
momentum $Q=k_{F}$ and consider the unitary limit. The impurity concentration
is $n_{\textrm{imp}}/n=0.15$.}
\end{figure}

In contrast, as we discussed earlier, the peak position nicely follows
the prediction $\omega_{\textrm{peak}}\simeq\hbar^{2}Q^{2}/(2m)-\mathcal{E}_{P}$
and are not sensitive to both temperature (apart from the temperature-dependence
in the polaron energy $\mathcal{E}_{P}$) and the impurity concentration.
Actually, the robustness of the peak position is one of the advantages
of the Raman spectroscopy mentioned in the experimental work \cite{Ness2020},
owing to the significant change in momentum. In Fig. \ref{fig4_ejRaman_nimp}
we check specifically the dependence of the Raman spectrum on the
impurity concentration $n_{\textrm{imp}}$. The spectrum becomes broader
with increasing $n_{\textrm{imp}}$, while keeps its peak position
nearly unchanged. In the inset, we subtract from $\omega_{\textrm{peak}}$
the background contribution $\hbar^{2}Q^{2}/(2m)$ and define $\omega_{0}=\omega_{\textrm{peak}}-\hbar^{2}Q^{2}/(2m)$.
We find that $\omega_{0}$ shows a very weak red-shift with increasing
impurity concentration at $Q=k_{F}$ (see the black dot-dashed line)
and correctly approaches $\left|\mathcal{E}_{P}\right|$ in the single-impurity
limit $n_{\textrm{imp}}\rightarrow0$. This can be contrasted with
the radio-frequency spectrum, where the peak position $\omega_{0}$
fails to recover the anticipated value $\left|\mathcal{E}_{P}\right|$
and there is a small systematic shift at about 0.02$\varepsilon_{F}$
in the single-impurity limit (i.e., the red dashed line). In Fig.
\ref{fig5_ejRaman_Tdep}, we also show the Raman spectrum at different
temperatures in the unitary limit. The spectrum becomes broader with
smaller peak height, as anticipated. 

\begin{figure}
\begin{centering}
\includegraphics[width=0.48\textwidth]{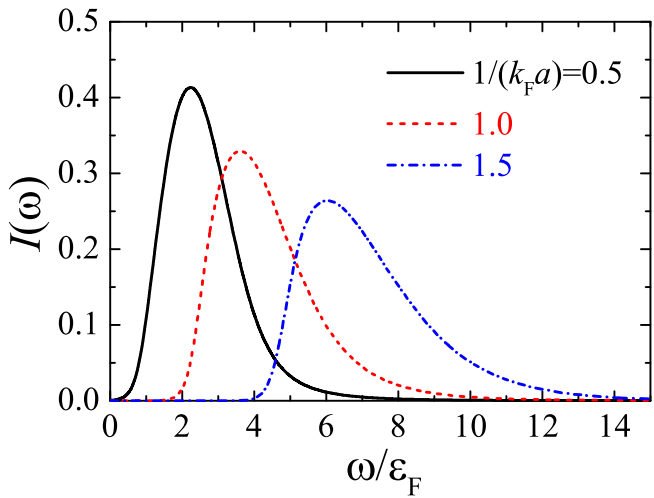}
\par\end{centering}
\caption{\label{fig6_ejRaman_vxdep} The ejection Raman spectrum $I(\omega$)
{[}in units of $\varepsilon_{F}^{-1}${]} below the Feshbach resonance
at three interaction strengths: $1/(k_{F}a)=0.5$ (black solid line),
$1.0$ (red dashed line), and $1.5$ (blue dot-dashed line). We take
a temperature $T=0.2T_{F}$ and a transferred momentum $Q=k_{F}$.
The impurity concentration is $n_{\textrm{imp}}/n=0.15$.}
\end{figure}

Finally, in Fig. \ref{fig6_ejRaman_vxdep} we report the ejection
Raman spectrum in the strong coupling regime with $1/(k_{F}a)>0$.
Our non-self-consistent $T$-matrix theory becomes less accurate when
we increase $1/(k_{F}a)$ at strong coupling \cite{Hu2022,Combescot2007}.
Nevertheless, we can see clearly that the spectrum becomes more and
more asymmetric with increasing $1/(k_{F}a)$. This is because the
coherent contribution from attractive polarons gets strongly suppressed.
The incoherent part from the molecule-hole continuum in the impurity
spectral function gives the major contribution to the Raman signal.
It then features a highly asymmetric energy tail in the spectrum above
the threshold $\omega_{\textrm{thres}}\simeq\hbar^{2}Q^{2}/(2m)+\mathcal{E}_{B}$,
where $\mathcal{E}_{B}$ is the binding energy of a molecule \cite{Ness2020}. 

\begin{figure}
\begin{centering}
\includegraphics[width=0.48\textwidth]{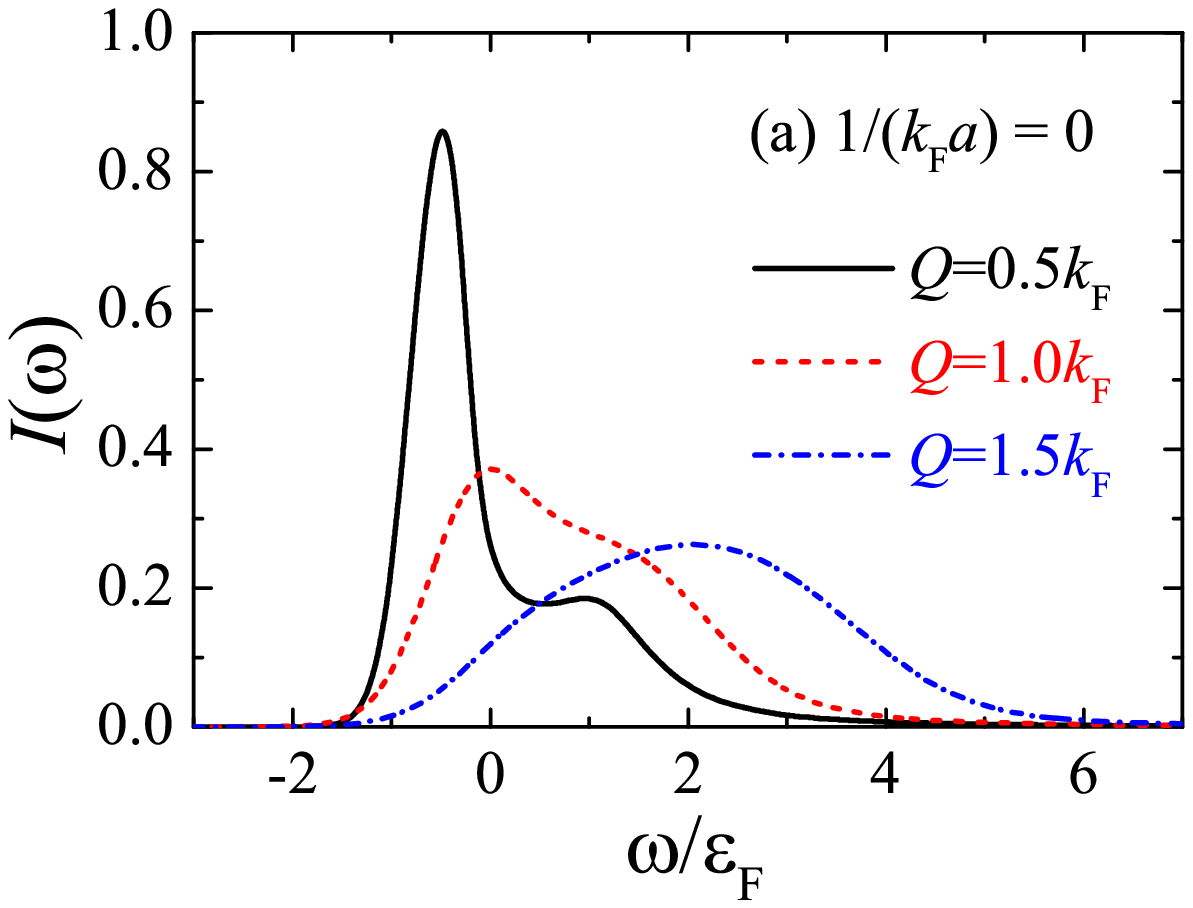}
\par\end{centering}
\begin{centering}
\includegraphics[width=0.48\textwidth]{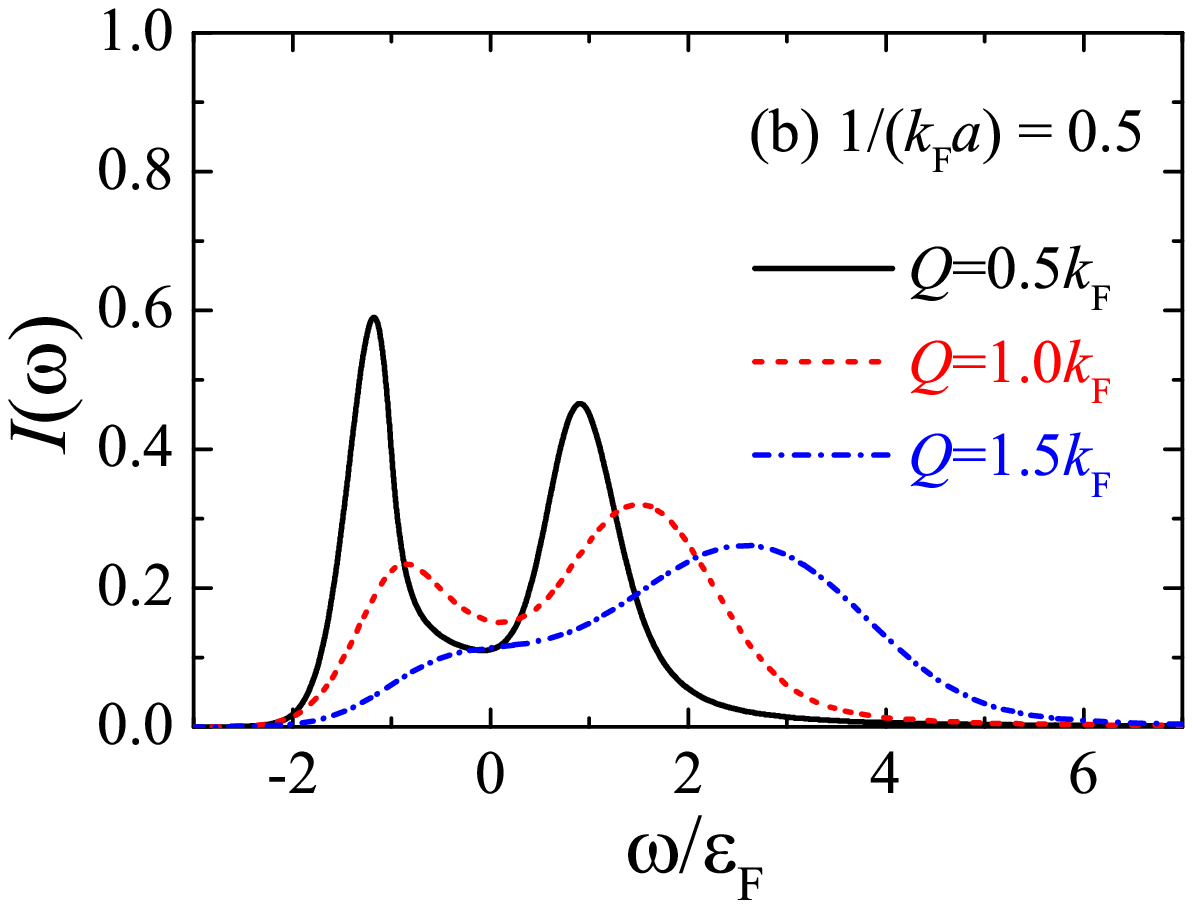}
\par\end{centering}
\caption{\label{fig7_injRaman} The injection Raman spectrum $I(\omega$) {[}in
units of $\varepsilon_{F}^{-1}${]} at the unitary limit $1/(k_{F}a)=0.0$
(a) and at $1/(k_{F}a)=0.5$ (b). We consider three different transferred
momenta as indicated in the plots. The temperature is $T=0.2T_{F}$
and the impurity concentration is $n_{\textrm{imp}}/n=0.15$. The
spectra are normalized to unity.}
\end{figure}

\subsection{Injection Raman spectroscopy }

We have so far discussed the ejection Raman spectroscopy, where the
contribution from the high-energy part of the single-particle spectral
function is thermally suppressed due to the Fermi distribution function
in Eq. (\ref{eq:ejectionRaman}). As a result, the excited state of
Fermi polarons, such as the repulsive polaron branch, can hardly be
probed in the ejection scheme. This problem can be solved, if the
impurity is initially prepared in a non-interacting state with the
Fermi bath. The Raman beams can then bring the impurity into the interacting
state, with a transfer rate as a function of the frequency recorded
as the spectrum. In this \emph{injection} scheme, the excited state
of Fermi polarons can be directly observed \cite{Scazza2017}. By
neglecting the initial-state effect, the injection Raman spectrum
at the transferred momentum $\mathbf{Q}$ is given by \cite{Torma2014,Parish2021},
\begin{equation}
I\left(\omega\right)=\frac{1}{V}\sum_{\mathbf{k}}A\left[\mathbf{k},\epsilon_{\mathbf{k}+\mathbf{Q}}^{(I)}+\omega\right]f\left(\epsilon_{\mathbf{k}+\mathbf{Q}}^{(I)}-\mu_{i}\right),
\end{equation}
where $\mu_{i}$ is the impurity chemical potential in the initial
non-interacting state, to be determined by the number equation, $n_{\textrm{imp}}=(1/V)\sum_{\mathbf{k}}f(\epsilon_{\mathbf{k}}^{(I)}-\mu_{i})$.

In Fig. \ref{fig7_injRaman}, we show the injection Raman spectrum
at the resonance (a) and at $1/(k_{F}a)=0.5$ (b). Three typical Raman
momenta are considered. In the unitary limit, for $Q\leq k_{F}$ we
see clearly the bump structure at the energy $\omega\sim\varepsilon_{F}$,
contributed from the precursor of the repulsive polaron. At $1/(k_{F}a)=0.5$,
the bumps develop into well-defined repulsive polaron peaks. In this
case, at $Q=0.5k_{F}$ the widths of the low-energy attractive polaron
peak and of the high-energy repulsive polaron peak are similar. Their
weights (i.e., the integrated area under each peak) are also similar.
However, as we increase the transferred momentum $Q$, the weight
gradually transfers to the repulsive polaron peak. At large transferred
momentum $Q=1.5k_{F}$, the contributions from attractive and repulsive
polaron basically merge into a very broad peak and somehow becomes
featureless. Therefore, we suggest that an optimized transferred momentum
$Q\sim k_{F}$ might be considered in the future experiments on the
injection Raman spectroscopy.

\begin{figure}
\begin{centering}
\includegraphics[width=0.5\textwidth]{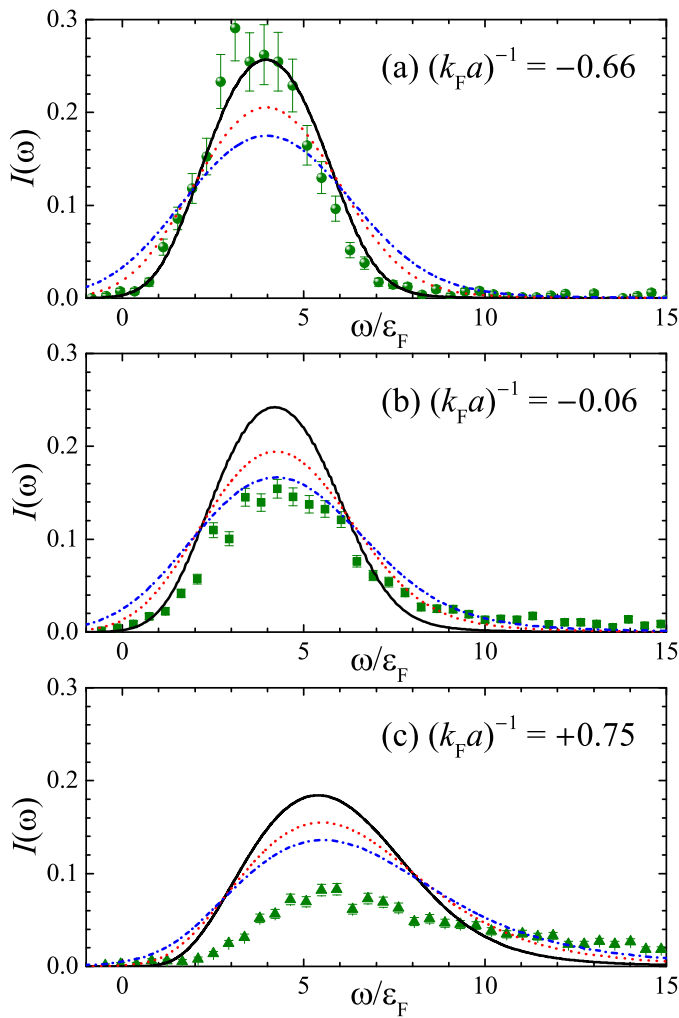}
\par\end{centering}
\caption{\label{fig8_expt} The comparison of the theory (lines) with the experimental
data from Ness \textit{et al.} (symbols) \cite{Ness2020}, for the
ejection Raman spectrum of a Fermi polaron across the Feshbach resonance.
The spectra are in units of $\varepsilon_{F}^{-1}$ and are normalized
to unity, $\int d\omega I(\omega)=1$. In the experiment, the temperature
of the background Fermi gas is $T=0.2T_{F}$ and the transferred momentum
is $Q=1.9k_{F}$. In our theoretical predictions, we consider three
different temperatures: $T=0.2T_{F}$ (black solid lines), $0.4T_{F}$
(red dashed lines) and $0.6T_{F}$ (blue dash-dotted lines). The impurity
density is taken as $n_{\textrm{imp}}/n=0.23$, following the experimental
condition after an average over the trap configuration \cite{Ness2020}.
In the comparison, we do not include any adjustable free parameters.}
\end{figure}

\subsection{Comparison with the experiment}

We are now in the position to compare our theory with the recent experiment
\cite{Ness2020}. As shown in Fig. \ref{fig8_expt}, we have considered
three different temperatures for the theoretical curves ($0.2T_{F}$,
$0.4T_{F}$, and $0.6T_{F}$, from the top to bottom), although experimentally
the Fermi bath temperature is about $0.2T_{F}$. The other parameters
such as the average impurity concentration $n_{\textrm{imp}}=0.23n$
and the interaction strength $1/(k_{F}a)$ follow the experimental
condition \cite{Ness2020}, so there are no free adjustable parameters.
The effect of spatial inhomogeneity caused by the external harmonic
traps is considered in Appendix, which does not lead to qualitatively
different results.

On the weak coupling side of the resonance (i.e., $1/(k_{F}a)=-0.66$
in Fig. \ref{fig8_expt}(a)), we find a good agreement between the
theoretical prediction and experimental data, both of which are given
at the temperature $0.2T_{F}$. This agreement is anticipated, since
our non-self-consistent $T$-matrix theory should work well with a
weak impurity-bath interaction.

However, near the Feshbach resonance (i.e., the middle plot Fig. \ref{fig8_expt}(b)
for the case $1/(k_{F}a)=-0.06$), our low temperature prediction
fails to explain the experimental data. A reasonable agreement for
the peak height might be reached, if we increase the Fermi bath temperature
to $0.6T_{F}$ in the theoretical calculation (see the blue dot-dashed
line). This temperature is significantly higher than the experimental
Fermi bath temperature. Tentatively, we attribute the disagreement
between theory and experiment to the inefficiency of the non-self-consistent
$T$-matrix theory for predicting the impurity spectral function,
although we do know that the theory predicts very accurate polaron
energy in the unitary limit at zero temperature \cite{Massignan2014,Combescot2007,Prokofev2008}.
On the other hand, it is worth mentioning that in the phenomenological
theory used in the experiment \cite{Ness2020}, the temperature of
the polaron $T_{P}$ extracted from fitting is also systematically
larger than the Fermi bath temperature ($0.2T_{F}$). Moreover, a
large background temperature $T_{\textrm{bg}}=2T_{F}$ is used for
the incoherent contribution from molecules. The larger polaron temperature
$T_{P}$ and background temperature $T_{\textrm{bg}}$ seem to be
consistent with our finding that a larger Fermi bath temperature is
theoretically needed to account for the experimental data.

At the strong coupling with $1/(k_{F}a)=0.75$ in Fig. \ref{fig8_expt}(c),
our non-self-consistent $T$-matrix theory is completely unable to
understand the experimental data. The data are about two times smaller
than our theoretical result at $0.2T_{F}$. The peak position read
from the data also seems to be larger than the theoretical prediction
(i.e., $\omega_{\textrm{peak}}\simeq\hbar^{2}Q^{2}/(2m)-\mathcal{E}_{P}$),
which we believe is reasonably accurate, since at zero temperature
the $T$-matrix theory predicts a polaron energy $\mathcal{E}_{P}$
that is in good agreement with quantum Monte Carlo simulation \cite{Massignan2014}.
We note that, the disagreement between theory and experiment at strong
coupling can hardly be reconciled by simply increasing the Fermi bath
temperature in the theoretical calculation. Our theoretical prediction
at $T=0.6T_{F}$ shown in the blue dot-dashed line is still far above
the experimental data.

\section{Conclusions and outlooks}

In summary, we have presented a microscopic calculation of Raman spectroscopy
of Fermi polarons at finite temperature, based on a well-documented
non-self-consistent many-body $T$-matrix theory \cite{Massignan2014,Hu2022,Combescot2007}.
The dependences of the ejection Raman spectrum on the transferred
Raman momentum, temperature, impurity concentration, and the impurity-bath
interaction are systematically investigated. We have also considered
the injection Raman spectrum, which might be experimentally measured
in the near future. 

We have compared our theoretical result with the recent ejection Raman
spectroscopy measurement \cite{Ness2020}. We have found a good agreement
in the weak coupling regime, which is encouraging and anticipated.
However, towards the Feshbach resonance and the strong coupling regime,
the non-self-consistent $T$-matrix theory cannot provide a quantitative
account of the experimental data. On the theoretical side, more refined
treatments are therefore needed. By extending the existing zero temperature
studies to the finite temperature case, the possible theoretical scenarios
could include the functional renormalization group \cite{Schmidt2011}
and the diagrammatic quantum Monte Carlo simulation \cite{Prokofev2008,Goulko2016}.
\begin{acknowledgments}
We thank Yoav Sagi and Richard Schmidt for fruitful discussions and
for sharing the experimental data with us. This research was supported
by the Australian Research Council's (ARC) Discovery Program, Grant
No. DP180102018 (X.-J.L).
\end{acknowledgments}

\appendix

\section{Raman spectrum within the local density approximation}

\begin{figure}
\begin{centering}
\includegraphics[width=0.5\textwidth]{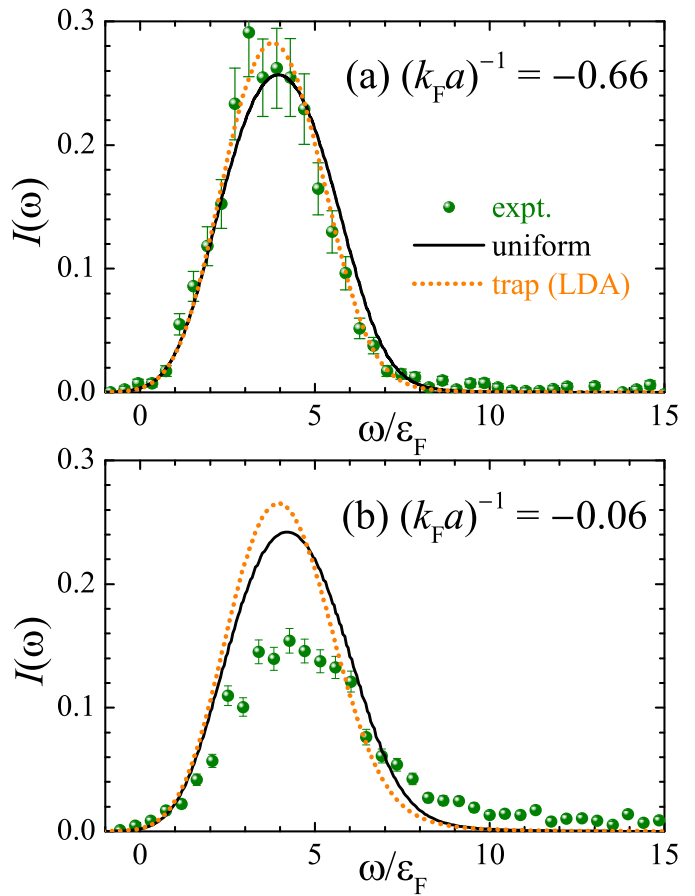}
\par\end{centering}
\caption{\label{fig9_exptLDA} The comparison of the theory (lines) with the
experimental data from Ness \textit{et al.} (symbols) \cite{Ness2020}
at the temperature $T=0.2T_{F}$ and transferred momentum $Q=1.9k_{F}$,
for the ejection Raman spectrum of a Fermi polaron on the BCS side
(a) and near the unitary limit (b). Here, $T_{F}=\varepsilon_{F}/k_{B}$
and $k_{F}$ are the peak Fermi temperature and Fermi wavevector of
majority fermions at the trap center, respectively. The spectra are
in units of $\varepsilon_{F}^{-1}$ and are normalized to unity, $\int d\omega I(\omega)=1$.
We present two theoretical predictions, for a uniform gas (black solid
line) and a trapped gas (orange dotted line), calculated without and
with the local density approximation, respectively. The local impurity
density is always $n_{\textrm{imp}}/n=0.23$.}
\end{figure}

In this Appendix, we consider the effect of spatial inhomogeneity
on the Raman spectrum, caused by the external harmonic traps. We assume
that the local density approximation (LDA) is applicable for the dynamical
quantities such as the Raman spectroscopy. As the ejection Raman spectrum
measures the transfer rate \emph{per impurity}, the trap-averaged
Raman spectrum takes the form \cite{Ness2020},
\begin{equation}
\left\langle I\left(\omega\right)\right\rangle =\frac{\int d\mathbf{r}I\left(\omega,\mathbf{r}\right)n_{I}\left(\mathbf{r}\right)}{\int d\mathbf{r}n_{I}\left(\mathbf{r}\right)},\label{eq:ejectionRFLDA}
\end{equation}
where $n_{I}(\mathbf{r})$ is the impurity density distribution and
$I\left(\omega,\mathbf{r}\right)$ is the ejection spectrum calculated
at the local position $\mathbf{r}$ with local majority fermion density
$n(\mathbf{r})$.

In general, the impurity distribution $n_{I}(\mathbf{r})$ is difficult
to obtain, due to the interaction with majority fermions that leads
to an (unknown) effective trapping potential. Here, since we are only
interested in a \emph{qualitative} estimate of the trapping effect,
we may assume a fixed ratio between the local impurity density to
the local majority fermion density, i.e., $n_{I}(r)/n(r)=n_{\textrm{imp}}/n=0.23$.
This assumption is reasonable, as the ejection Raman spectrum seems
to depend weakly on $n_{\textrm{imp}}/n$, as shown in Fig. \ref{fig4_ejRaman_nimp}.
Therefore, we replace in Eq. (\ref{eq:ejectionRFLDA}) $n_{I}(\mathbf{r})$
by $n(\mathbf{r})$.

The theoretical trap-averaged ejection Raman spectra at $1/(k_{F}a)=-0.66$
and $1/(k_{F}a)=-0.06$ are shown in Fig. \ref{fig9_exptLDA}, in
the form of the orange dotted lines. As expected, the trap-average
does not show a qualitative difference, in comparison with the theoretical
predictions for a uniform gas, which have been discussed in detail
in the main text. However, for the interaction parameter $1/(k_{F}a)=-0.66$
in Fig. \ref{fig9_exptLDA}(a), the trap-average does lead to a better
agreement between theory and experiment, particularly at large frequency
(i.e., $\omega>5\varepsilon_{F}$). The disagreement between theory
and experiment found near the unitary limit or on the strong-coupling
BEC side can not be resolved by taking the trap-average.

\end{document}